# On a minimum Column Density for Massive Star Formation

*C Sivaram*

Indian Institute of Astrophysics, Bangalore, 560 034, India

Telephone: +91-80-2553 0672; Fax: +91-80-2553 4043

e-mail: sivaram@iiap.res.in

*Kenath Arun*

Christ Junior College, Bangalore, 560 029, India

Telephone: +91-80-4012 9292; Fax: +91-80- 4012 9222

e-mail: kenath.arun@cjc.christcollege.edu

**Abstract:** A fundamental physical basis is provided for the recently obtained minimum column density (1gcm$^{-2}$) for massive star formation.

In a recent letter [1], it was shown that only clouds with column densities of at least 1gcm$^{-2}$ can avoid fragmentation and form massive stars. Efficient radiative cooling keeps most collapsing star-forming clouds close to isothermal and this results in fragmentation into stars of one solar mass or lower. Cloud heating can prevent this fragmentation and allow massive star formation.

Here we point out that this threshold column density of 1gcm$^{-2}$ can be simply understood in terms of physics underlying radiation pressure and gravity. We can invoke in this context that for a given mass M (which may be constituted of stars, gas etc.) there is a maximal luminosity (so called Eddington luminosity) given by:

$$L = \frac{4\pi G M m_P c}{\sigma_T} \qquad \ldots (1)$$

where $\sigma_T = \frac{8\pi}{3}\left(\frac{e^2}{m_e c^2}\right)^2$, $m_P$, $c$ are the Thomson cross-section, proton mass and the velocity of light.

The radiation pressure force acting on the system then gives rise to an acceleration acting outwards away from the cloud, i.e., tending to dissipate it, which is given by:

$$a_{Rad} = \frac{Gm_P}{\sigma_T} \qquad \ldots (2)$$

(Note, the radiative acceleration is simply given as $\approx \frac{L}{Mc}$)

This dissipative outward acceleration must be balanced by an inward gravitational acceleration given by:

$$a_{Grav} = \frac{GM}{R^2} \qquad \ldots (3)$$

The requirement $a_{Grav} \geq a_{Rad}$ then gives the threshold column density of:

$$\frac{M}{R^2} \approx \frac{m_P}{\sigma_T} \sim 1 g/cm^{-2} \qquad \ldots (4)$$

This accounts for the fact that we need a minimum column density of 1gcm$^{-2}$ (expressed entirely in terms of the fundamental constants following from physics) of:

$\frac{M}{R^2} \approx \frac{m_P}{\sigma_T} \sim 1 g/cm^{-2}$ (as $\sigma_T$ & $m_P$ are both of the same numerical order, i.e. $10^{-24} cm^{-2}$ & $10^{-24} g$).

A lower column density would give a lesser gravitational effect to hold the system against breaking up.

It is remarkable that in an entirely different context [2 - 6], we had obtained a threshold mass surface density of 1gcm$^{-2}$ to form large scale structures in the universe wherein here the balance is between the repulsive dark energy density (e.g. cosmological constant) and the self gravity energy density. This implied:

$$\frac{GM^2}{8\pi R^4} \approx \frac{\Lambda c^2}{8\pi G} \qquad \ldots (5)$$

Implying $\frac{M}{R^2} \approx \sqrt{\Lambda} \frac{c^2}{G} \sim 1 g/cm^{-2} \qquad \ldots (6)$

(for the observed value of $\Lambda \approx 10^{-56} cm^{-2}$ from the current dark energy formation)

Indeed all structures, ranging from galaxies to super-clusters seem to have this observed surface mass density value of 1gcm$^{-2}$. [7, 8]

So in an entirely different context we have this intriguing value. Here this is the minimum column density to form cosmic structures, as given by equation (6).